# Tracking of individual freely diffusing fluorescent protein molecules in the bacterial cytoplasm

Brian P. English†[1], Arash Sanamrad†[1], Stoyan Tankov[1, 2], Vasili Hauryliuk[1, 2] & Johan Elf[1*]

[1]Department of Cell and Molecular Biology, Uppsala University, Uppsala, Sweden
[2]University of Tartu, Institute of Technology, Tartu, Estonia

†Equal contribution
*johan.elf@icm.uu.se

ABSTRACT

We combine stroboscopic laser excitation with stochastic photoactivation and super-resolution fluorescence imaging. This makes it possible to record hundreds of diffusion trajectories of small protein molecules in single bacterial cells with millisecond time resolution and sub-diffraction limited spatial precision. We conclude that the small protein mEos2 exhibits normal diffusion in the bacterial cytoplasm with a diffusion coefficient of $13.1 \pm 1.2\ \mu m^2\ s^{-1}$. This investigation lays the groundwork for studying single-molecule binding and dissociation events for a wide range of intracellular processes.

INTRODUCTION

Our intuitive grasp of the molecular mechanisms of life is centered around the behavior of an individual molecule: we perceive its static structure from crystallographic models, we envision its dynamic nature and we outline kinetic schemes that show how it interacts with other molecules. However, we almost never observe this molecule directly, and our mental construction of it is mainly an inference from experiments performed with Avogadro numbers of molecules. Technological advances have made biological investigations of single-molecule interactions possible (van Oijen, 2008). From individual trajectories we can extract additional information such as dynamic or static intrinsic heterogeneity, since it is not hidden by ensemble averaging, and thus intermediate states are readily revealed (Lu *et al.*, 1998, Zhuang *et al.*, 2002, English *et al.*, 2006). However, the vast majority of these experiments are only conducted in test tubes, not in living cells.

What kind of data is needed for a direct assessment of biochemical interactions *in vivo*? The criteria are defined by the spatial and temporal characteristics of the biological system. Sub-diffraction limited spatial resolution (in the nanometer range) is necessary for an adequate structural description, and high time resolution (in the millisecond range) combined with high temporal dynamic range (spanning milliseconds to seconds) is needed for an adequate temporal analysis.

Here we introduce an *in vivo* single-molecule assay that meets all these criteria by combining recent advances in single-molecule imaging and tracking. The single-molecule tracking approach readily meets the requirement for sub-diffraction limited spatial resolution (Deich *et al.*, 2004, Schmidt *et al.*, 1996). While current imaging technology allows for observation of slow processes, it potentially biases the observed distribution towards slower subpopulations. In the worst case scenario, fast millisecond dynamics may be completely overlooked. We circumvent this limitation by combining this imaging technique with a technique borrowed from high-speed photography: stroboscopic illumination (Yu *et al.*, 2006, Elf *et al.*, 2007).

Our fluorescent probe is the green-to-red photoconvertible GFP variant mEos2 (McKinney *et al.*, 2009) which we stochastically convert to re-spawn single fluorophores. mEos2 expands our biological observation window to high copy targets since we can control the number of bright re-spawned molecules via tunable activation pulses (Betzig *et al.*, 2006, Rust *et al.*, 2006).

To test the validity of our method, we have chosen the hardest biological target to track, a small 26-kDa protein molecule. In doing so we lay the groundwork for investigations of all larger and thus slower biological systems. Single-molecule tracking of small protein molecules has scientific value per se: the nature of how they diffuse in the crowded bacterial cytoplasm is a matter of debate for decades already, and a variety of diffusion and sub-diffusion models have been put forward (Dix & Verkman, 2008, Magdziarz *et al.*, 2009). Sub-diffusion has been observed for large complexes in the bacterial cytoplasm (Golding & Cox, 2006) as well as for molecules in cell membranes (Ghosh & Webb, 1994). If the movement of small protein molecules in the cytoplasm were also sub-diffusive it would profoundly affect the validity of test-tube experiments as *in vivo* models.

We record hundreds of trajectories per cell, which allows for accurate determination of local apparent diffusion coefficients throughout the cell. By analyzing this extensive dataset and by taking into account confinement effects we establish that small protein molecules exhibit normal diffusion in the bacterial cytoplasm at least down to the 4-ms timescale.

## MATERIALS AND METHODS

### LASER MICROSCOPY

The optical setup is shown schematically in Figure 1a. An acousto-optical modulator (AOM, IntraAction, 40 MHz) shutters a wide-field yellow excitation laser beam into an Olympus IX81 inverted microscope. This 555-nm DPSS laser (CrystaLaser) is collimated by an Olympus TIRF objective (NA = 1.45) and excites an area with a 2-µm radius at the sample plane with a laser power of 50 to 200 $kWcm^{-2}$. The AOM is synchronized with a PhotonMax EMCCD camera (Princeton Instruments) by a NI-DAQ M-series data acquisition card (PCI-6259, National Instruments), and is controlled via LabVIEW 8.5 to pass short 0.4- or 1-ms excitation pulses in the middle of each imaging frame. A photoactivation laser beam at 405 nm (Radius, Coherent) is spatially overlapped using a long-pass dichroic filter (Z405RDC, Chroma), and is focused at the sample plane. The photoactivation laser is independently shuttered using a UNIBLITZ T132 shutter controller that delivers 3-ms activation pulses at regular time intervals at powers ranging from 0.1 $kWcm^{-2}$ to 40 $MWcm^{-2}$. A long-pass dichroic filter (Z555RDC, Chroma) is used to excite the cells with both the excitation and photoactivation lasers and also to separate the emission light. *E. coli* cells (MG1655) expressing the monomeric fluorescent protein mEos2 (McKinney *et al.*, 2009) from a plasmid derived from the pDendra2-B vector (Evrogen) are imaged on a M9-glucose agarose pad in a FCS2 flow chamber (Bioptechs).

The duration of the excitation pulse is a critical parameter for obtaining distinct fluorescence spots from the rapidly moving single molecules. The optimal value is a tradeoff between collecting the large numbers of photons needed to determine the position of the molecule accurately and the broadening of the spot due to diffusion of the molecule during the excitation pulse. The optimal value under our experimental conditions was determined to be in the range of 0.4 to 1 ms.

We can convert the photoactivation laser from wide-field excitation to confocal excitation mode via a flip-lens (see Figure 1a). This allows us to record reverse fluorescence recovery after photobleaching (reverse FRAP) or photoactivation (PA) ensemble data to complement our single-molecule analysis for the same individual *E. coli* cells (see Results).

Individual *E. coli* cells are imaged for 20000 to 60000 frames at 250 Hz using up to 28 pixel lines on a PhotonMax EMCCD camera. The photoactivation frequency and laser power are adjusted for each cell to have on average less than one fluorophore visible at any given time. Example frames are shown in Figure 1b. The time-lapse movies are then edited by hand to remove frames with photoactivation pulses and frames where we observe multiple molecules in one frame.

## ANALYSIS AND SIMULATIONS

We track the fluorophores in our hand-edited movies using the particle tracking software Diatrack (v3.03, Semasopht), which identifies and fits the intensity spots of our fluorescent particles with symmetric 2D Gaussian functions (Figure 1b). In Figure 1c we show two experimental trajectories of individual, freely diffusing mEos2 molecules in an individual *E. coli* cell.

A standard analysis of a particles's diffusion trajectory is the calculation of its mean square displacements (MSDs, $\langle \Delta x^2 \rangle$) at different time intervals. In a system without spatial confinement, the MSD of a diffusing particle is proportional to time to the power of $\alpha$, i.e. $\langle \Delta x^2 \rangle \propto t^\alpha$. $\alpha$ is equal to unity for normal diffusion, larger than unity for super-diffusive particles and less than unity in the sub-diffusion case. While the interpretation of MSDs are straightforward for freely diffusing molecules, e.g. in the test-tube scenario, confinement effects by the geometry of the cell must be considered *in vivo*. Hence the experimental analysis must be complemented by simulations of diffusion in the given geometry of each cell (Deich *et al.*, 2004, Niu & Yu, 2008), and in our analysis we follow suit (see below).

Another standard approach is to calculate the cumulative distribution functions (CDFs) of displacements. This approach is a straightforward way to obtain apparent diffusion coefficients by fitting them to an exponential function corresponding to 2D Brownian motion (see Figure 3a) (Schutz *et al.*, 1997, Vrljic *et al.*, 2002).

We implement both approaches. All routines for trajectory analyses are written in IGOR Pro 6.12A. For each of the analyzed cells all trajectories are first overlaid to determine the precise cellular geometry. This geometry is then approximated as a cylinder with two hemispherical end caps. The final geometry of the cell is determined by subtracting twice the standard deviation of the estimated fitting noise (40 nm) from all sides of the initial geometry of the cell. This corrects for apparent cell broadening due to the experimental fitting noise. Only points that are inside the new geometry are included in the analysis. All experimentally obtained trajectories containing more than three entries are kept and their coordinates are transformed such that the cell's long axis is aligned with the x-axis.

MSDs along both axes (Figure 3b) and MSDs along the long axis of the cell (Figure 4a and b) are calculated for all possible time intervals. Standard errors of the means (SEMs) are calculated from the standard deviations of the square displacements and the number of contributing trajectories.

Since we collect hundreds of trajectories, diffusing molecules sample the entire cell. This allows us to calculate the geometry of the cytosol by compensating for the experimental fitting error, which increases the observed volume of the cell. The local apparent diffusion coefficients displayed in Figure 5a and b are evaluated every 20 by 20 nm in an x-y grid. The apparent diffusion coefficient assigned to a point in the grid is calculated from the mean square displacement along the long axis of the cell at 4, 8, 12 and 16 ms for all displacements originating within a radius of 200 nm from this point.

Simulations are performed to determine whether our trajectories can be well described by normal diffusion. The simulations consist of 3D random walks sampled at the 10-μs timescale in cells with the experimentally obtained geometries. The starting points of the trajectories are sampled from a uniform distribution. The positions of the points of the trajectories are calculated by averaging the positions of the corresponding points during the experimental exposure times.

The trajectory lengths are determined probabilistically. Each trajectory is terminated with a probability $1 - kp(x)$ every simulated frame (4 ms), where $p(x)$ is the probability density function to find a molecule at position $x$ along the long axis. $p(x)$ is determined from the distribution of starting points of the experimentally observed trajectories. $p(x)$ is included in the termination condition to correct for any unevenness in the illumination of the cell. $k$ is a scale factor chosen such that the average number of frames in a trajectory is approximately equal for experimental and simulated trajectories.

We add two types of noise to the simulated trajectories. First we add Gaussian-distributed movement noise. The standard deviation of this noise is given by

$$\sigma^2 = 2D\tau, \quad (1)$$

where $D$ is the diffusion coefficient and $\tau$ is the exposure time. This is to account for the uncertainty in the position of the molecule, which arises from the movement of the molecule during the exposure time. The movement noise is re-sampled for each point until the point is inside the geometry of the cell. We then add Gaussian-distributed fitting noise, which arises from the limited number of photons that are detected from the molecule during the exposure time.

We obtain 95% confidence intervals for the MSD plots in Figure 4a and b by calculating and sorting MSDs for 1000 simulations with the same number of trajectories as in the experiments.

To analyze the photoactivation experiments, the experimental data is first averaged over four consecutive photoactivation events each encompassing 30 frames taken over 120 ms. The pre-activation background is subtracted from each frame and the experimental intensities are projected on the long axis of the cell resulting in the experimental intensity distribution. The fluorescence intensity is modeled by the 1D diffusion equation with reflecting boundaries. The equation is integrated from the initial condition given by the fluorescence intensity distribution in the first camera frame after photoactivation. The diffusion coefficient is optimized such that the model error is minimized. Here photo-bleaching in the experimental data is corrected for by a correcting factor. The error norm is only minimized over the central part of the cell, since the polar regions have a disproportionally small intensity when projected to 1D.

## RESULTS

We have obtained experimental diffusion trajectories from eight individual *E. coli* cells on M9-based agarose pads at room temperature. In Figure 2a, we show all 1354 analyzed trajectories obtained from one of these cells. We use stroboscopic illumination to localize the fluorescence spots of freely diffusing protein molecules. Our stroboscopic illumination time is 400 μs for two cells and 1 ms for the remaining cells. Our fitting precision is proportional to the square root of the number of detected photons (Thompson *et al.*, 2002). While we could easily freeze any movement within a diffraction-limited spot using pulses shorter than 200 μs, precise localization is not possible given the poor photo-physical properties of mEos2 and hence we image at higher illumination pulse durations. The frames are recorded at the highest read-out speed of our EMCCD camera (4 ms for 28 pixel lines). We can record at lower frame rates, but this makes consecutive positions nearly uncorrelated and increases the risk of assigning molecules to wrong trajectories. Trajectory lifetimes are exponentially distributed with an average of 5 frames (see Figure 2b).

We analyze our single-molecule diffusion trajectories by calculating CDFs of displacements (Schutz *et al.*, 1997, Vrljic *et al.*, 2002) and MSDs (Deich *et al.*, 2004, Niu & Yu, 2008) in the x-y plane as depicted in Figure 3a and b, respectively. For all our eight cells, we observe cell-averaged apparent diffusion coefficients of $8.1 \pm 1.0\ \mu\text{m}^2\ \text{s}^{-1}$ for CDFs and $8.9 \pm 0.9\ \mu\text{m}^2\ \text{s}^{-1}$ for MSDs (both are for 4-ms steps) (see Table 1). These apparent diffusion coefficients, while being quite consistent among all cells analyzed, are not microscopic diffusion coefficients. This is all the more apparent when the first two steps (4 and 8 ms) of the MSDs are analyzed. Then we observe a much lower apparent diffusion coefficient of $4.0 \pm 1.0\ \mu\text{m}^2\ \text{s}^{-1}$. A strong correlation between cell size and apparent diffusion coefficients is further evidence that we are not observing microscopic diffusion coefficients (see Table 4).

It is clear that the microscopic diffusion coefficients have to be determined using another method. Confinement effects are less severe along the long axis of cells. We therefore construct MSDs along the long axis of cells to separate diffusion across cells in which our molecules instantaneously interact with the cell wall from less confined diffusion along the long axis.

In Figure 4a and b we show experimental MSDs along the long axis obtained from two individual *E. coli* cells. Analysis along the long axis results in much larger apparent diffusion coefficients as compared to analysis of trajectories in the x-y plane (see Table 1) as the confinement effects are less severe along the long axis. However, since the coefficients are still correlated with the size of the cells some confinement effects are still present (see Table 2). To accurately account for these confinement effects and to obtain microscopic diffusion coefficients ($D_{\mathrm{micro}}$), we simulate trajectories assuming normal diffusion in the volumes defined by the geometries of the cells. As can be seen in Figure 4a and b, the simulated confidence intervals perfectly describe our experimental data, which strongly indicates that the small mEos2 protein indeed performs simple Brownian motion, i.e. normal diffusion, at longer timescales than 4 ms. The same analysis was repeated for six additional cells of varying cell sizes. The results are summarized in Table 3. Unlike the apparent diffusion coefficients, $D_{\mathrm{micro}}$ does not vary with cell size, and a more uniform $D_{\mathrm{micro}}$ for all *E. coli* cells regardless of cell size is now apparent. The Pearson's correlation coefficient for correlation between cell length and $D_{\mathrm{micro}}$ shows statistical absence of correlation ($r = -0.02$) (see Table 4).

Having established that our molecule displays normal diffusion in all of our analyzed cells, we will now analyze the spatial distribution of apparent diffusion coefficients due to geometric constraints within an individual cell. The large number of experimental trajectories for each cell obtained with sub-diffraction limited resolution allows us to dissect how the apparent diffusion coefficients change throughout the cell. Figure 5a displays the spatial distribution of apparent diffusion coefficients along the long axis and Figure 5b displays the distribution of apparent diffusion coefficients in the x-y plane for the same cell as in Figure 4a. There appears to be a large variation in diffusion coefficients across the cell. This variation is especially noticeable in the faster timescales. This variation can be erroneously interpreted as variations in the intracellular viscosity. This non-uniformity is fully described by confinement effects since identical patterns emerge when we simulate normal diffusion within a uniform cell with the same number of trajectories as obtained in the experiments (see Figure 5a and b, right panel). This corroborates that the entire cytoplasm obeys normal diffusion. The large spatial variation in the apparent diffusion coefficient is due to the geometry constraint and the fitting inaccuracy for each point in the trajectories. The geometry makes apparent diffusion faster in the middle of the cells where molecules can diffuse less restricted for 4 ms and slower in the quarter positions where molecules have encountered the wall and returned, giving a lower apparent diffusion coefficient.

We independently obtain microscopic diffusion coefficients from ensemble photoactivation (PA) experiments performed on two of the previously analyzed cells (see Figure 6). When we use PA to determine the diffusion coefficients of mEos2 under our experimental conditions, our best estimate of $D_{\mathrm{micro}}$ is 11 $\mu m^2\ s^{-1}$. However, this estimate is dependent on the assumed geometry of the individual cell as was noted already by Elowitz *et al.* (Elowitz *et al.*, 1999). However, the main conceptual problem with ensemble PA experiments is that one has to establish normal diffusion before one can obtain microscopic diffusion coefficients at all.

Our analysis shows that the fluorescent proteins display normal diffusion in the cell at least down to the 4-ms timescale. This implies that the small protein perceives the *E. coli* cytoplasm very differently from the large RNA-protein complexes that previously have been demonstrated to display sub-diffusion when tracked at the single-molecule level (Golding & Cox, 2006). Despite the excellent agreement with a normal diffusion model at the greater than 4-ms timescale, our MSD curves do not extrapolate to zero at time zero. Instead we observe a pronounced offset of $\sim 0.05\ \mu m^2$. At first glance this offset might be attributed to sub-diffusive behavior at timescales faster than 4 ms. However, experimental fitting noise can cause pseudo-sub-diffusive behavior as any experimental fitting error will shift the MSD curves to higher values and therefore contribute to any offset (Thompson *et al.*, 2002).

## DISCUSSION

We present for the first time microscopic characterization of the diffusion process for a small freely diffusing cytoplasmic protein molecule in individual growing cells at room temperature under well-controlled physiological conditions. This is achieved by combining stroboscopic illumination with photoactivation.

The photoactivation approach is much more controllable than fine-tuning of GFP induction, which is non-linear in inducer concentration (Choi *et al.*, 2008). Furthermore, photoactivation can be modulated at much higher time resolution (milliseconds vs. typically several hours due to slow maturation rates of current dyes) and photoactivation pulses can be applied multiple times during the same experiment. This sequential activation and tracking of individual fusion proteins has previously been used to characterize slowly diffusing targets in bacterial cells (Biteen *et al.*, 2008, Niu & Yu, 2008). We now combine this approach with stroboscopic excitation, creating a setup which is capable of following fast biological processes.

We contrast our method with traditional methods for studying diffusion in living cells. Diffusion in individual cells has been successfully studied using ensembles of molecules such as fluorescence recovery after photobleaching (FRAP) or photoactivation (PA) (Elowitz et al., 1999, Mullineaux *et al.*, 2006, Konopka *et al.*, 2009) and fluorescence correlation spectroscopy (FCS) (Edman *et al.*, 1999, Magde *et al.*, 1972).

In the case of FRAP or PA a large number of fluorescent molecules are bleached or activated in a part of the cell and the spread of molecules into the dark areas are monitored over time. Not only do such techniques require an abundance of fluorescent molecules, limiting them to high copy number targets, they are also intrinsically limited in spatial resolution by the diffraction limit that restricts the analysis to the coarsest length scales found in bacteria. This coarseness fundamentally also limits temporal resolution in such assays.

Figure 6 illustrates that it is possible to obtain accurate diffusion coefficients using a PA assay on the same cells as were studied using our single-molecule assay. However, it should be pointed out that interpreting FRAP or PA data is highly geometry dependent which could in the worst case scenario be misinterpreted as either cell to cell variability (due to different geometries between cells) or spatial heterogeneity within one cell.

Here we explicitly simulate normal diffusion of photoconverted proteins along the long axis of the cell given the cell's geometry, the photoactivation profile and the bleaching characteristics. This allows us to reproduce the experimental PA data and hence we obtain microscopic diffusion coefficients. One caveat is that normal diffusion has to be established first by an independent method (as we have done) before this method can be used to obtain such coefficients as this method would not readily reveal deviations from Brownian motion. Also, such ensemble methods are much less sensitive to heterogeneity, be it static (different populations) or dynamic (molecules switching between states, such as binding or dissociation), and studies of intracellular binding kinetics with PA is therefore likely out of reach.

Another well-established technique, FCS, has excellent temporal resolution but its temporal dynamic range is restricted to narrow diffusion times through a diffraction-limited focal volume. Furthermore, while FCS is a good choice for low abundant cytosolic biomolecules, any slow kinetics may be misinterpreted as static heterogeneity.

Our method heavily relies on a very large data set, which we were able to obtain using expression from high-copy number plasmids and by regenerating the fluorescence using repeated photoactivation cycles. For an accurate determination of microscopic diffusion coefficients hundreds of individual trajectories per cell are needed.

This may not be readily available for a chromosomally encoded GFP-tagged biological target of choice. First, the copy number of the target may be much lower as the majority of bacterial proteins are expressed at very low copy numbers (Xie *et al.*, 2008). Second, the photo-chemistry of GFP variants are both pH and redox sensitive (Bogdanov *et al.*, 2009a, Bogdanov *et al.*, 2009b), thus limiting the number of observable molecules *in vivo* even further. We have also observed much dimmer mEos2 fluorescence in the cytoplasm, and were able to recover *in vitro* photo-physics by burning small holes into the cell membrane with the activation beam in a confocal configuration and at powers of 4 $MWcm^{-2}$ (data not shown).

If hundreds of trajectories per cell are not achievable, apparent diffusion coefficients are still readily obtainable for only a few trajectories per cell from the CDFs and MSDs, even without taking cell geometries into consideration. While in this case we are not able to obtain microscopic diffusion coefficients anymore, nevertheless apparent diffusion coefficients are still reproducible between cells and directly correspond to microscopic diffusion coefficients for slow-moving molecules when the MSD is linear within the millisecond time range.

In conclusion, we have demonstrated the possibility to characterize intracellular diffusion at high temporal precision and spatial resolution by using single-protein tracking in the cytoplasm of living bacterial cells. We have reliably tracked the protein mEos2 in the cytosol, and we established that its diffusion characteristics are indistinguishable from Brownian motion. This inherent intracellular inertness, coupled with a time resolution on par with the fastest cellular protein diffusion dynamics, makes this method a general tool for investigation of intracellular kinetics. This will make it possible to use genetic fusions to mEos2 and similar GFP variants to study binding of our molecule of choice to its binding partners where deviations from Brownian diffusion will be biologically interesting.

In the near future this presented methodology will be employed for studying single-molecule binding and dissociation events *in vivo* for wide range of intracellular processes.

TABLES

**Table 1. Apparent diffusion coefficients along both axes of the cell ($D_{x\text{-}y}$).**

| Cell number | Exposure time (ms) | Frame time (ms) | Cell length (μm) | Apparent $D_{x\text{-}y}$ (CDF) ($\mu m^2\ s^{-1}$) | Apparent $D_{x\text{-}y}$ (MSD, 4 ms) ($\mu m^2\ s^{-1}$) | Apparent $D_{x\text{-}y}$ (MSD, 4 and 8 ms) ($\mu m^2\ s^{-1}$) |
|---|---|---|---|---|---|---|
| 1 | 0.4 | 4 | 1.7 | 7.7 | 8.6 | 3.2 |
| 2 | 0.4 | 4 | 2.6 | 9.1 | 10.1 | 5.5 |
| 3 | 1 | 4 | 2.2 | 7.8 | 9.1 | 3.5 |
| 4 | 1 | 4 | 2.0 | 6.9 | 7.9 | 3.9 |
| 5 | 1 | 4 | 3.0 | 9.8 | 10.3 | 5.1 |
| 6 | 1 | 4 | 1.8 | 8.2 | 8.9 | 3.2 |
| 7 | 1 | 4 | 2.5 | 7.7 | 8.5 | 4.7 |
| 8 | 1 | 4 | 1.9 | 7.2 | 8.0 | 2.7 |
| Average | | | 2.2 | 8.1 | 8.9 | 4.0 |
| Standard deviation | | | 0.5 | 1.0 | 0.9 | 1.0 |

**Table 2. Apparent diffusion coefficients along the long axis of the cell ($D_x$).**

| Cell number | Exposure time (ms) | Frame time (ms) | Cell length (μm) | Apparent $D_x$ (CDF) ($\mu m^2\ s^{-1}$) | Apparent $D_x$ (MSD, 4 ms) ($\mu m^2\ s^{-1}$) | Apparent $D_x$ (MSD, 4 and 8 ms) ($\mu m^2\ s^{-1}$) |
|---|---|---|---|---|---|---|
| 1 | 0.4 | 4 | 1.7 | 7.9 | 13.2 | 5.1 |
| 2 | 0.4 | 4 | 2.6 | 10.5 | 15.6 | 9.9 |
| 3 | 1 | 4 | 2.2 | 10.7 | 14.8 | 6.2 |
| 4 | 1 | 4 | 2.0 | 13.2 | 12.7 | 7.0 |
| 5 | 1 | 4 | 3.0 | 13.8 | 15.3 | 9.1 |
| 6 | 1 | 4 | 1.8 | 8.3 | 12.9 | 5.1 |
| 7 | 1 | 4 | 2.5 | 17.2 | 12.6 | 8.0 |
| 8 | 1 | 4 | 1.9 | 7.1 | 11.3 | 4.5 |
| Average | | | 2.2 | 11.1 | 13.6 | 6.9 |
| Standard Deviation | | | 0.5 | 3.5 | 1.5 | 2.0 |

**Table 3. Inferred microscopic diffusion coefficients ($D_{\mathrm{micro}}$).** The microscopic diffusion coefficients are in bold.

| Cell number | Exposure time (ms) | Frame time (ms) | Cell length (μm) | Scale factor for PDF | $D_{\mathrm{micro}}$ ($\mu m^2\ s^{-1}$) |
|---|---|---|---|---|---|
| 1 | 0.4 | 4 | 1.7 | 0.80 | **13** |
| 2 | 0.4 | 4 | 2.6 | 1.50 | **14** |
| 3 | 1 | 4 | 2.2 | 1.05 | **13** |
| 4 | 1 | 4 | 2.0 | 0.85 | **15.5** |
| 5 | 1 | 4 | 3.0 | 1.85 | **12.5** |
| 6 | 1 | 4 | 1.8 | 0.90 | **11.5** |
| 7 | 1 | 4 | 2.5 | 1.10 | **12.5** |
| 8 | 1 | 4 | 1.9 | 1.00 | **13** |
| Average | | | 2.2 | | **13.1** |
| Standard deviation | | | 0.5 | | **1.2** |

**Table 4. Pearson's correlation coefficients for diffusion coefficients and cell lengths.** The correlation coefficient for $D_{\text{micro}}$ is in bold.

| Method | $\boldsymbol{D_{\text{micro}}}$ | Apparent $D_x$ (CDF) | Apparent $D_x$ (MSD, 4 ms) | Apparent $D_x$ (MSD, 4 and 8 ms) | Apparent $D_{x\text{-}y}$ (CDF) | Apparent $D_{x\text{-}y}$ (MSD, 4 ms) | Apparent $D_{x\text{-}y}$ (MSD, 4 and 8 ms) |
|---|---|---|---|---|---|---|---|
| Cell length | $\boldsymbol{-0.02}$ | 0.66 | 0.67 | 0.89 | 0.76 | 0.75 | 0.88 |

FIGURES

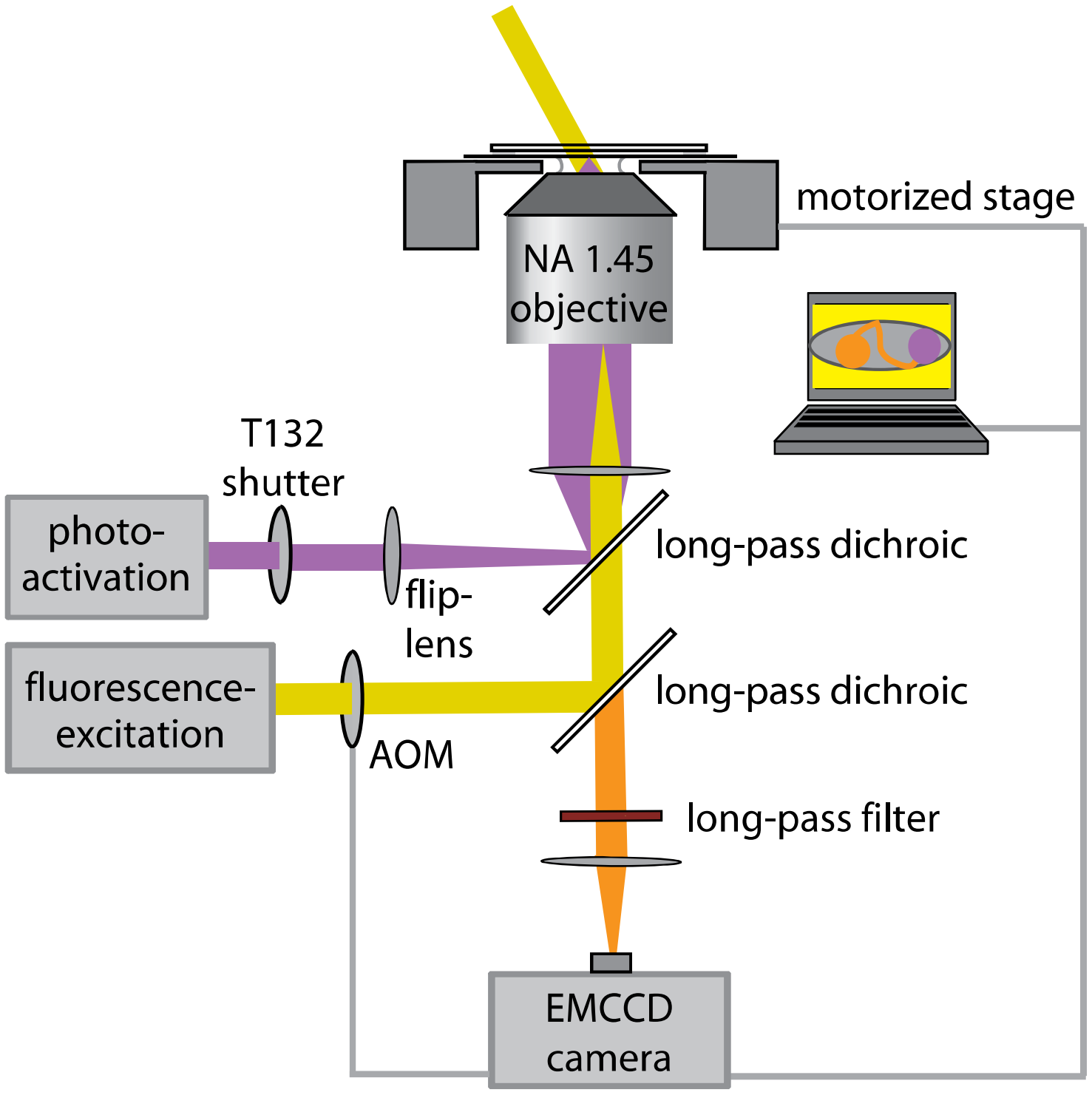

**Figure 1a. Schematic diagram of the optical setup.**

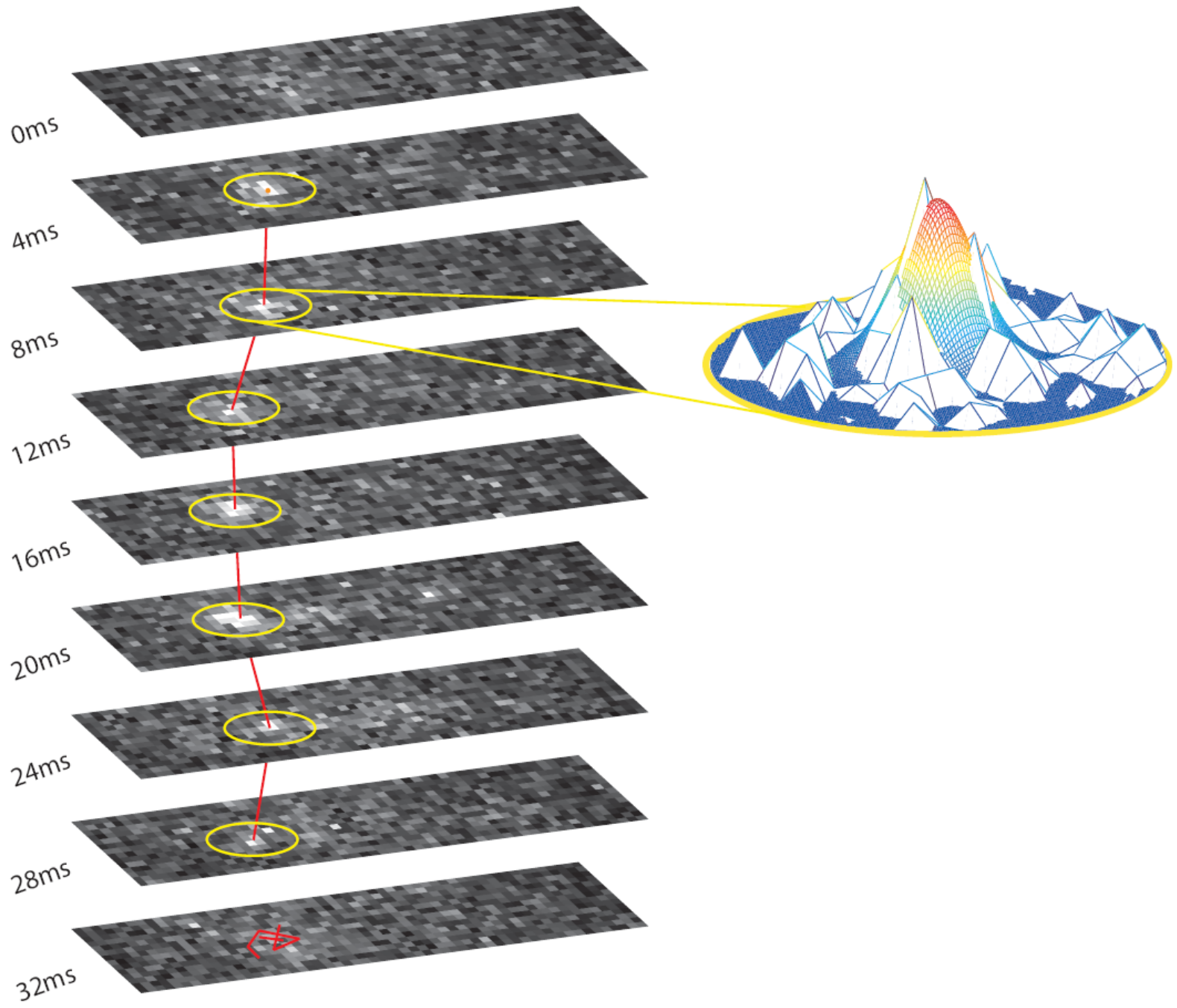

**Figure 1b. Nine consecutive frames of a single mEos2 trajectory tracking and a Gaussian fit to frame 3.**

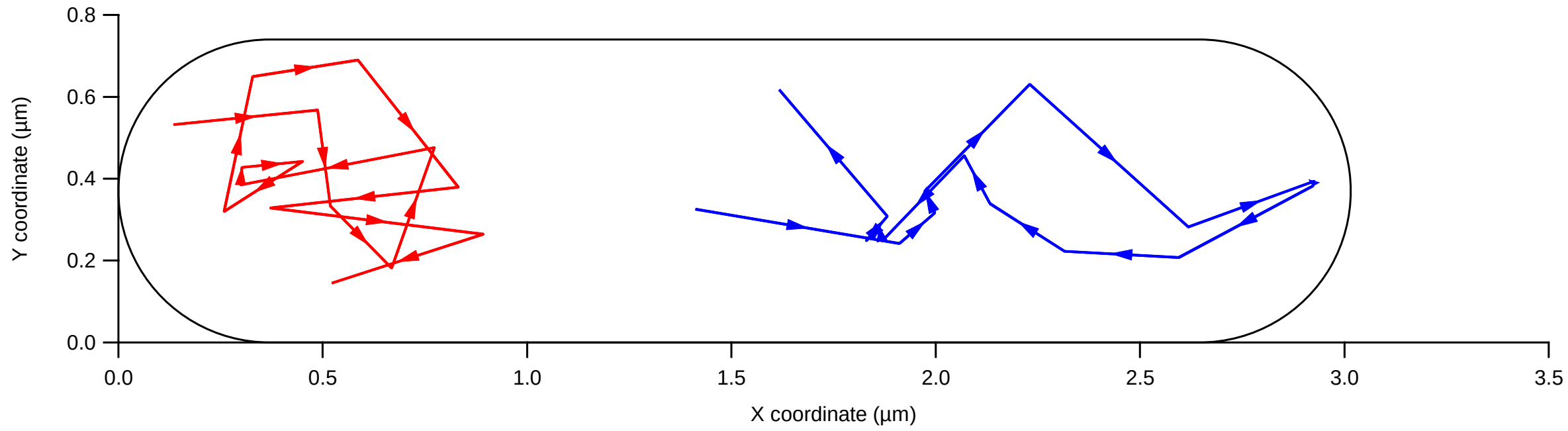

**Figure 1c. Overlay of two experimentally obtained single-molecule mEos2 trajectories from an *E. coli* cell.**

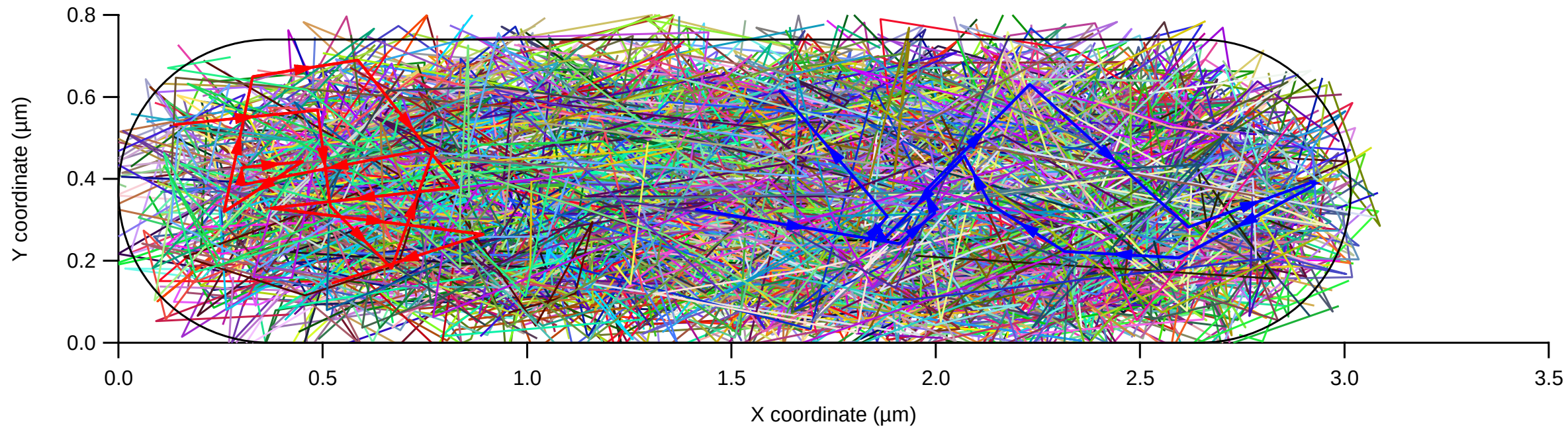

**Figure 2a. Overlay of all 1354 single-molecule trajectories for *E. coli* cell number 5 (see Table 1 and 2).**

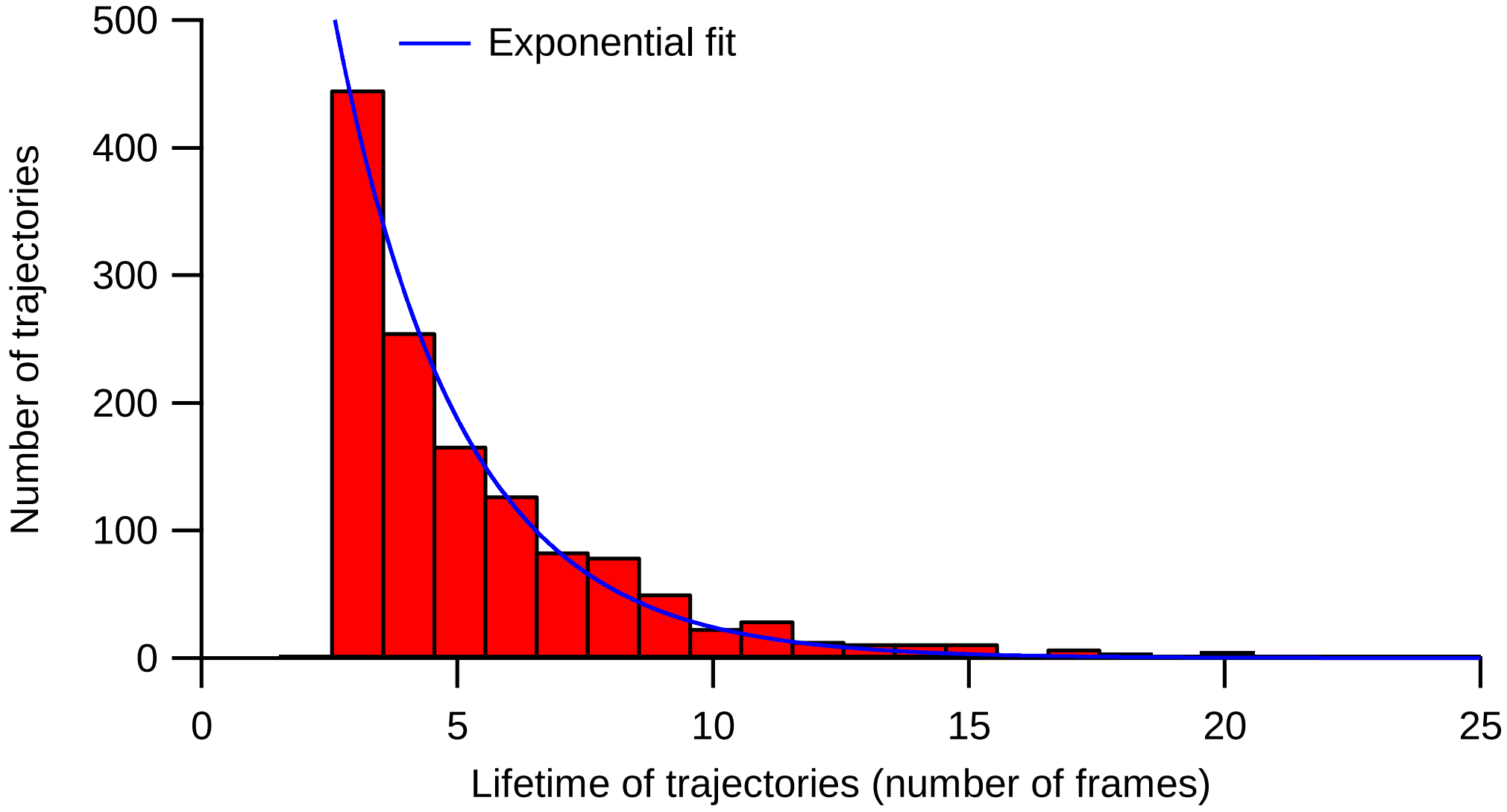

**Figure 2b. Histogram of the trajectory lifetimes for the 1354 analyzed trajectories presented in Figure 2a.**

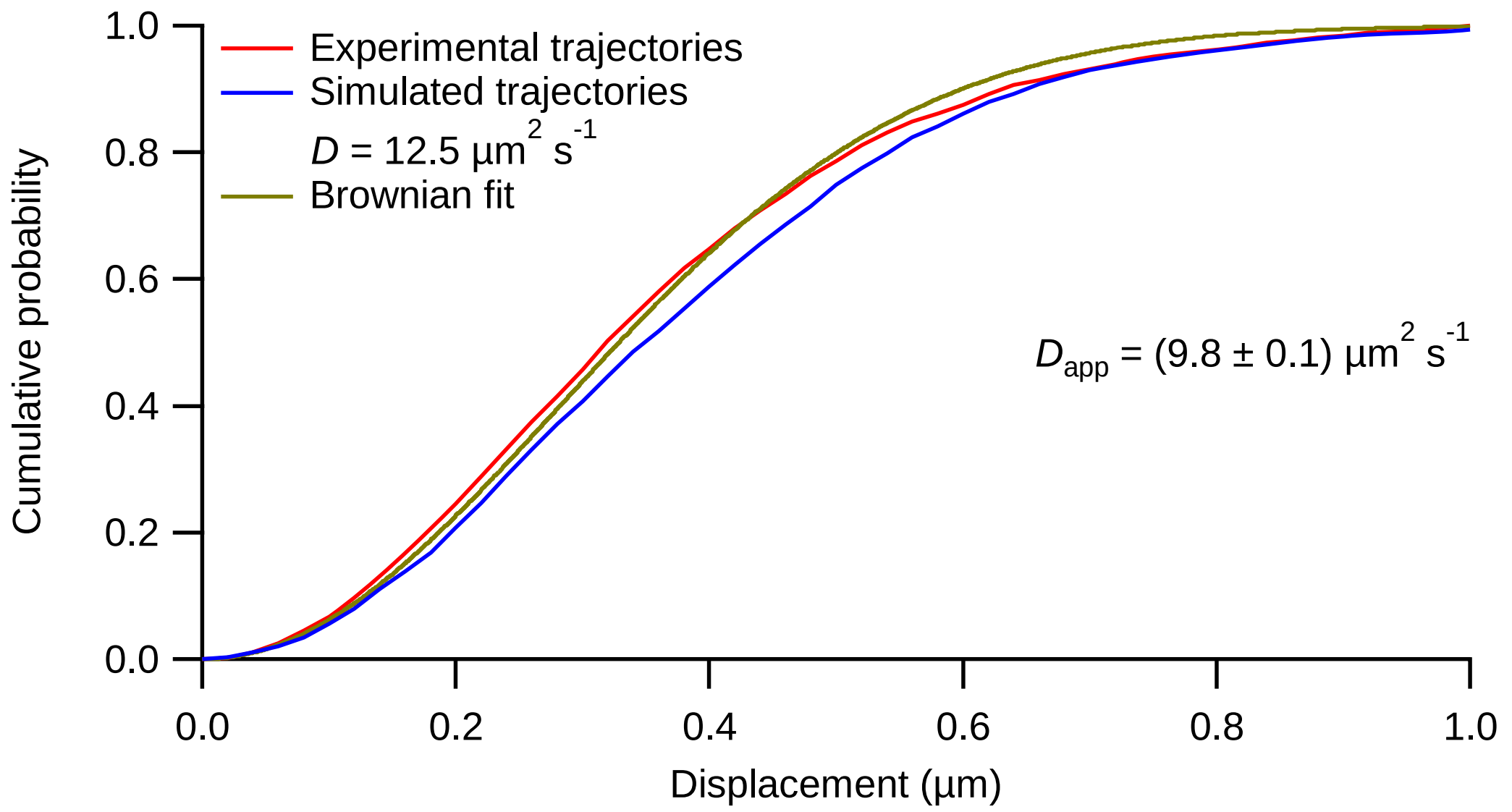

**Figure 3a. Cumulative distribution function (CDF) of the displacements *r* along the x- and y-axes of an *E. coli* cell.** We obtain an apparent diffusion coefficient of $9.8 \pm 0.1\ \mu m^2\ s^{-1}$ by fitting $P(r, \Delta t) = 1 - e^{-\frac{r^2}{4D_{\text{app}}\Delta t}}$ (gold) to the experimental CDF (in red). The CDF of simulated trajectories in the geometry of the cell is also shown (blue).

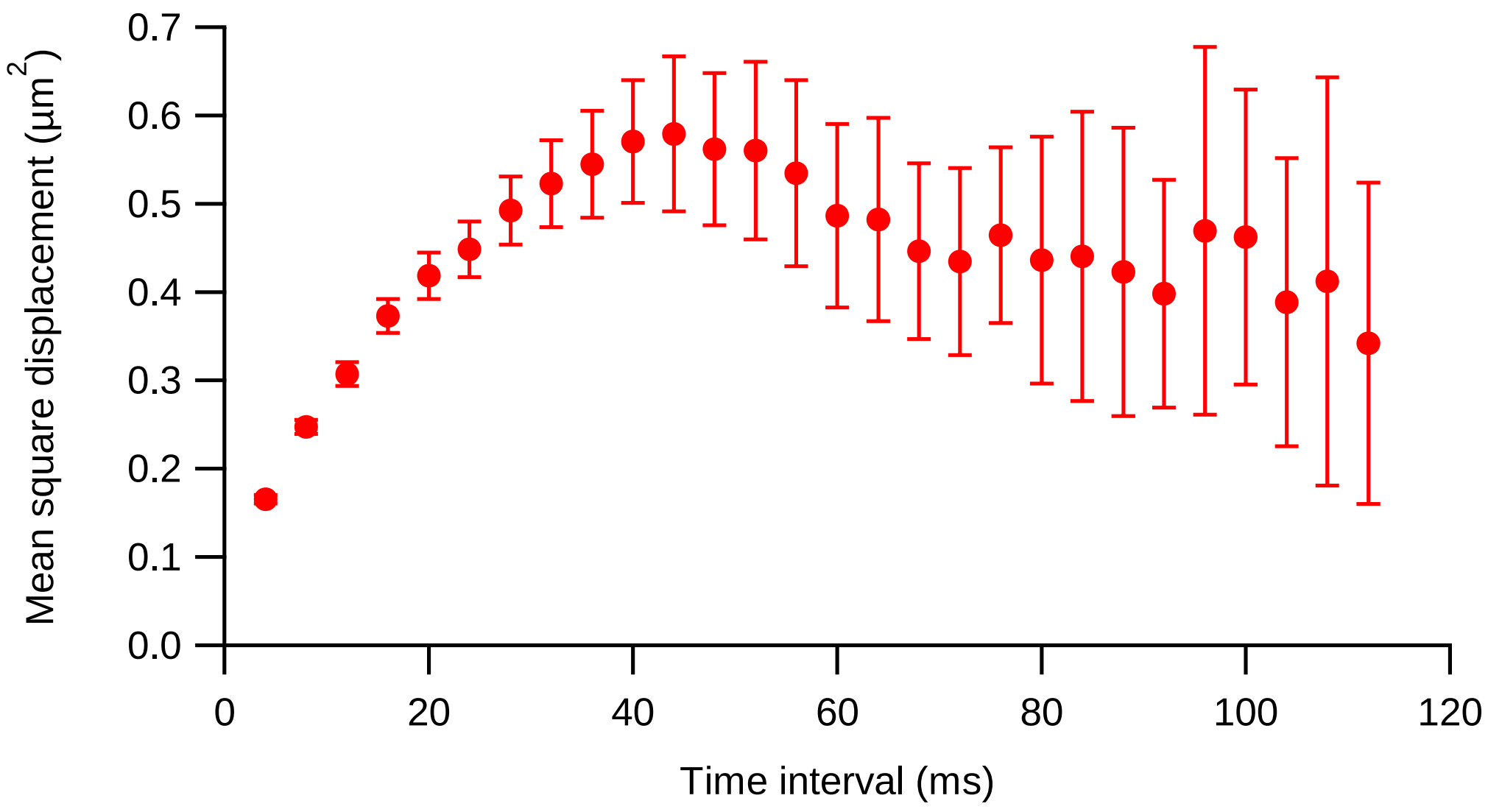

**Figure 3b. Mean square displacements (MSDs) along the x- and y-axes of an *E. coli* cell for different time intervals.** The error bars represent the experimental standard errors of the means.

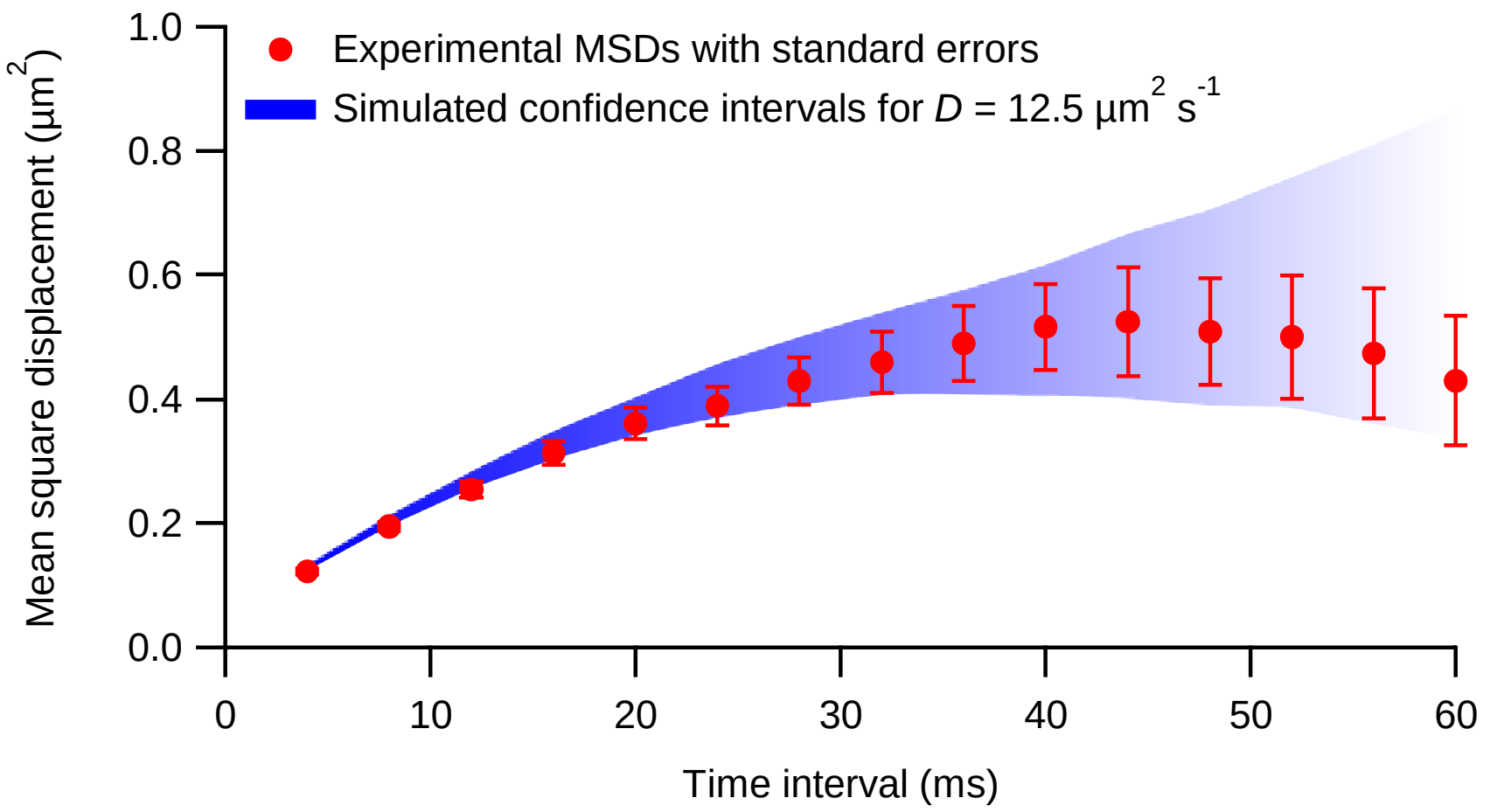

**Figure 4a. Mean square displacements (MSDs) along the long axis of *E. coli* cell 5 for different time intervals.** The error bars represent experimental standard errors of the means. The 95% confidence intervals of simulated trajectories are displayed in blue for normal diffusion at $D = 12.5\ \mu m^2\ s^{-1}$ in the volume defined by the geometry of this cell. The simulations account for the effects of experimental fitting noise, the unevenness in the excitation illumination, and the lifetime distribution of the experimental trajectories (see Methods).

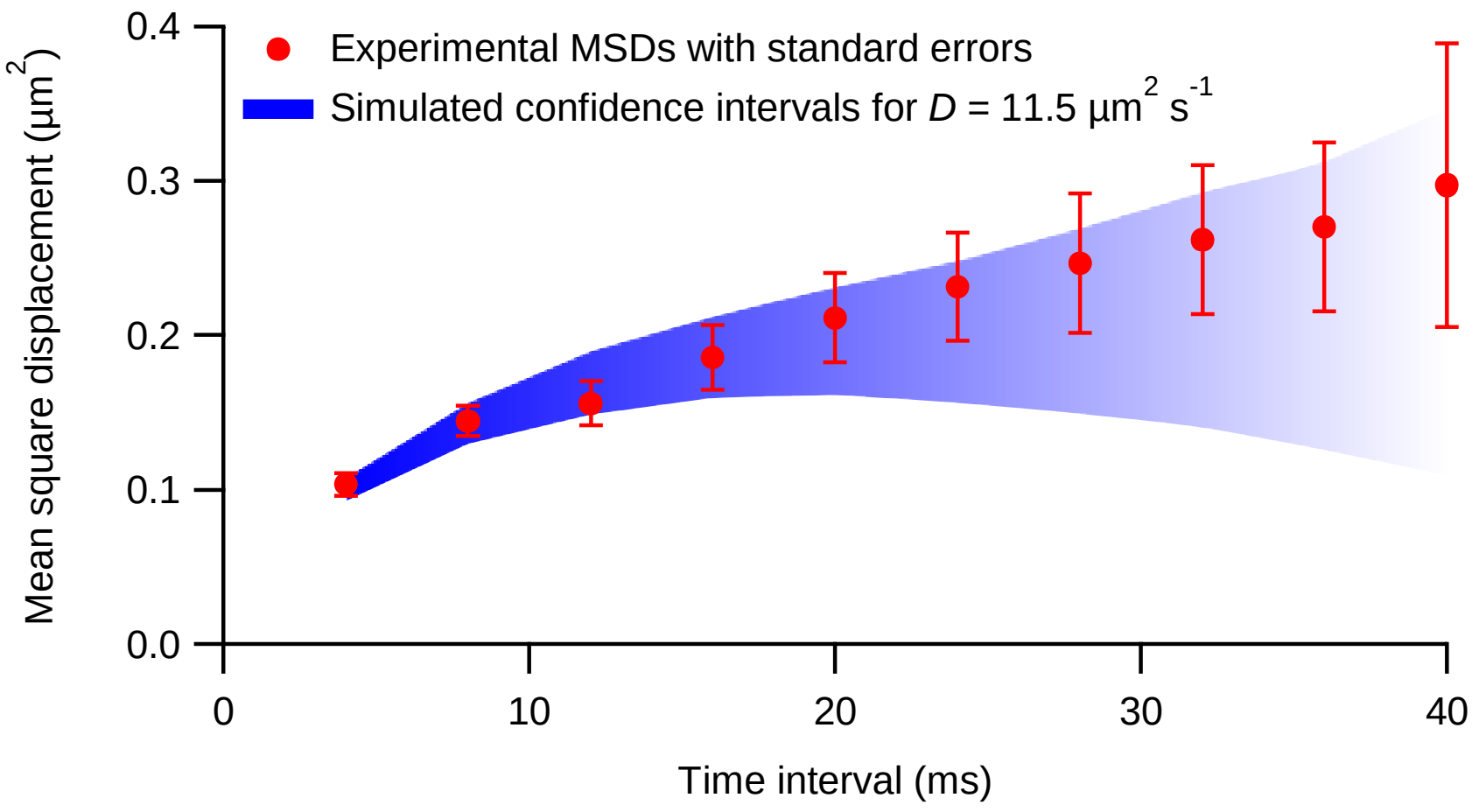

**Figure 4b. Mean square displacements (MSDs) along the long axis of *E. coli* cell 6 for different time intervals.** The error bars and confidence intervals were calculated as in figure 4a for normal diffusion at $D = 11.5\ \mu m^2\ s^{-1}$.

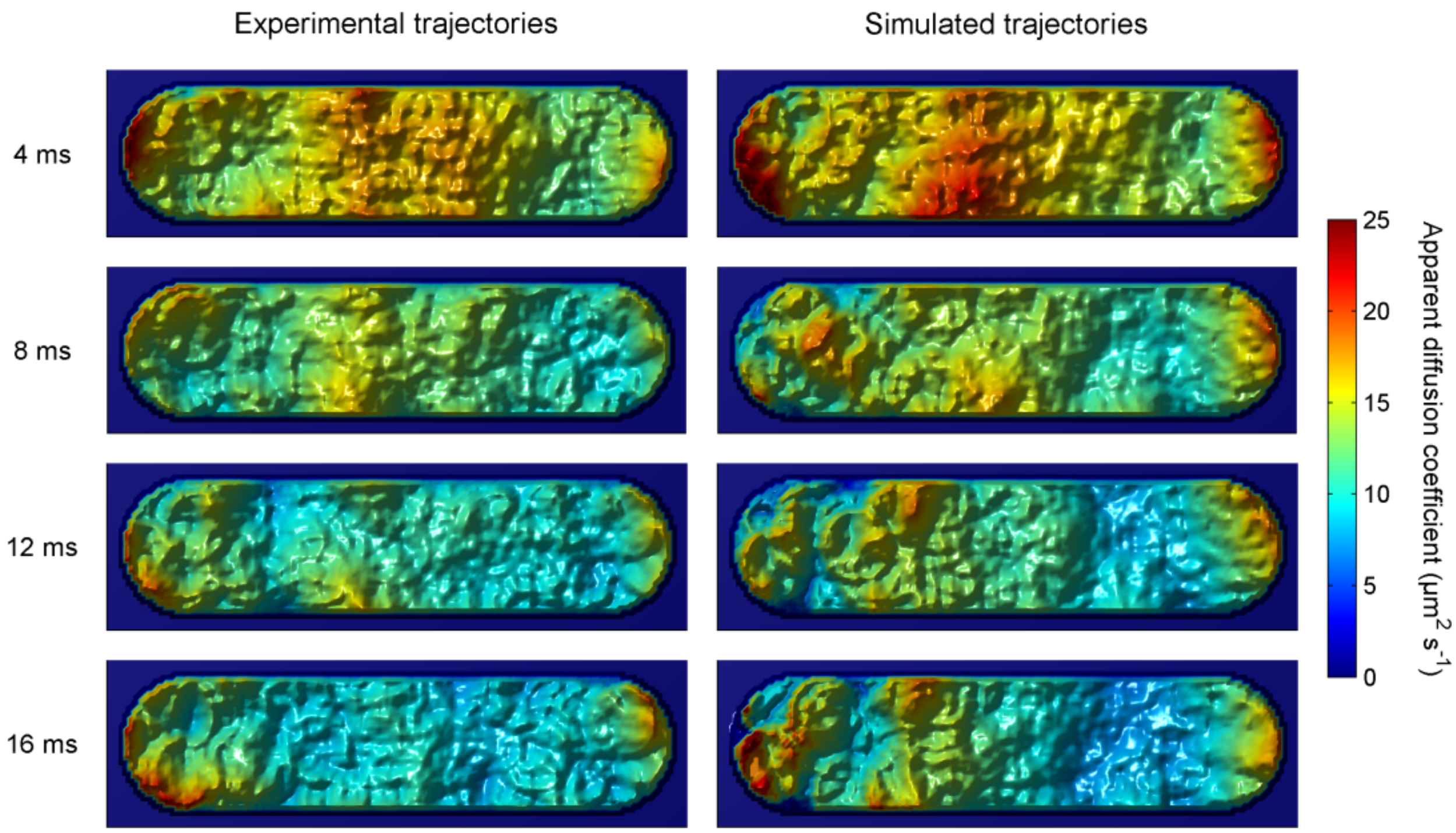

**Figure 5a. Local apparent diffusion coefficients along the long axis of *E. coli* cell 5.** Each point in the graph is false-colored according to the mean square displacement calculated over 4, 8, 12 and 16 ms for experimental (left panel) and simulated (right panel) displacements originating in a circle ($r = 200$ nm). The simulations assume normal diffusion at $D = 12.5\ \mu m^2\ s^{-1}$ in the volume defined by the geometry of this cell.

The apparent diffusion coefficients are higher in the middle of the cell, as the molecules can diffuse more freely along the long axis, especially at shorter timescales. We can also see in both panels that noise contributions make the apparent diffusion faster close to the cell wall.

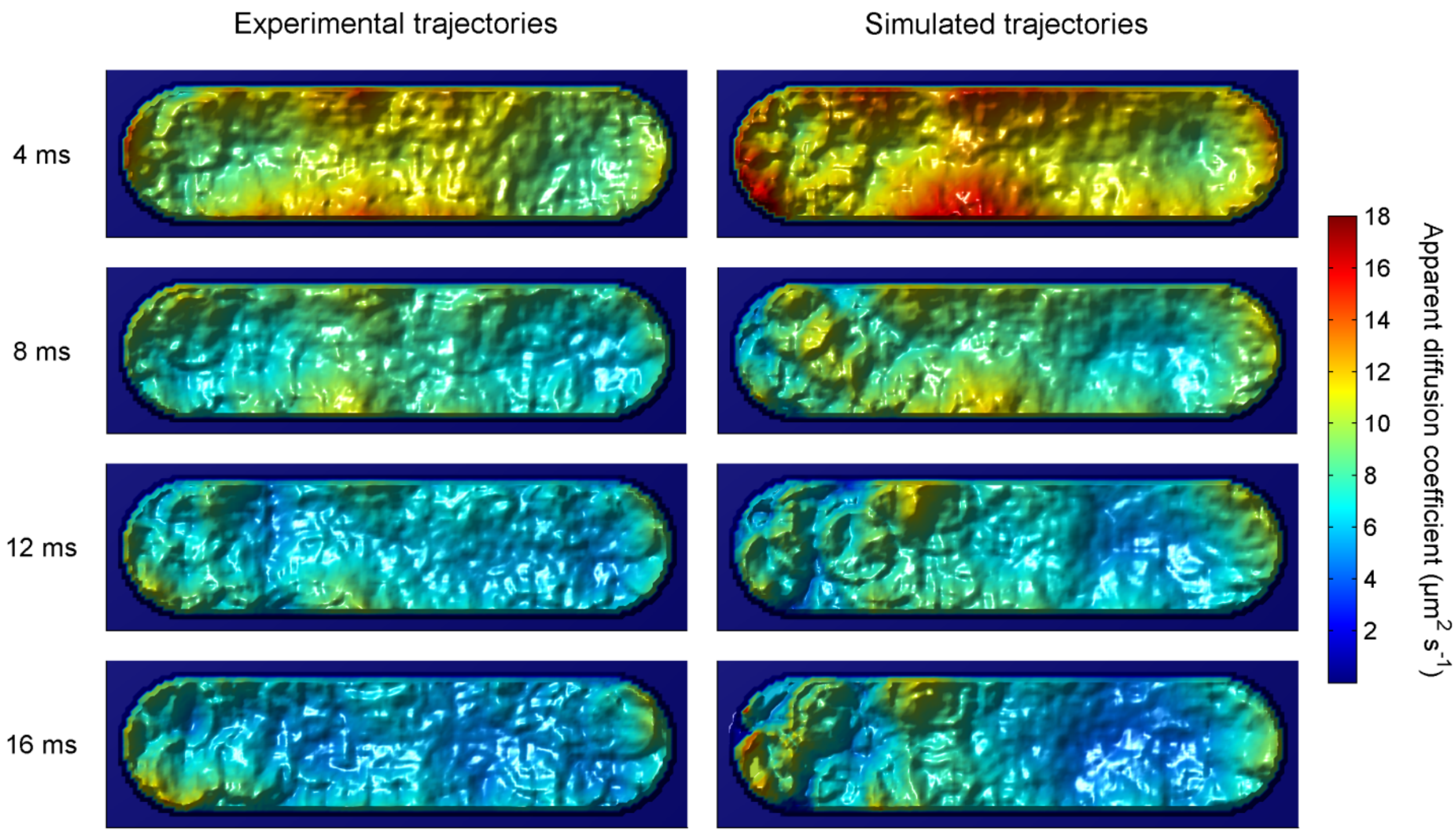

**Figure 5b. Local apparent diffusion coefficients along both axes of *E. coli* cell 5.** Each point in the graph is false-colored according to the mean square displacement calculated over 4, 8, 12 and 16 ms for experimental (left panel) and simulated (right panel) displacements originating in a circle ($r = 200$ nm). The simulations assume normal diffusion at $D = 12.5\ \mu\text{m}^2\ \text{s}^{-1}$ in the volume defined by the geometry of this cell.

As also seen in figure 5a, there is good agreement between experimental and simulated local apparent diffusion coefficients.

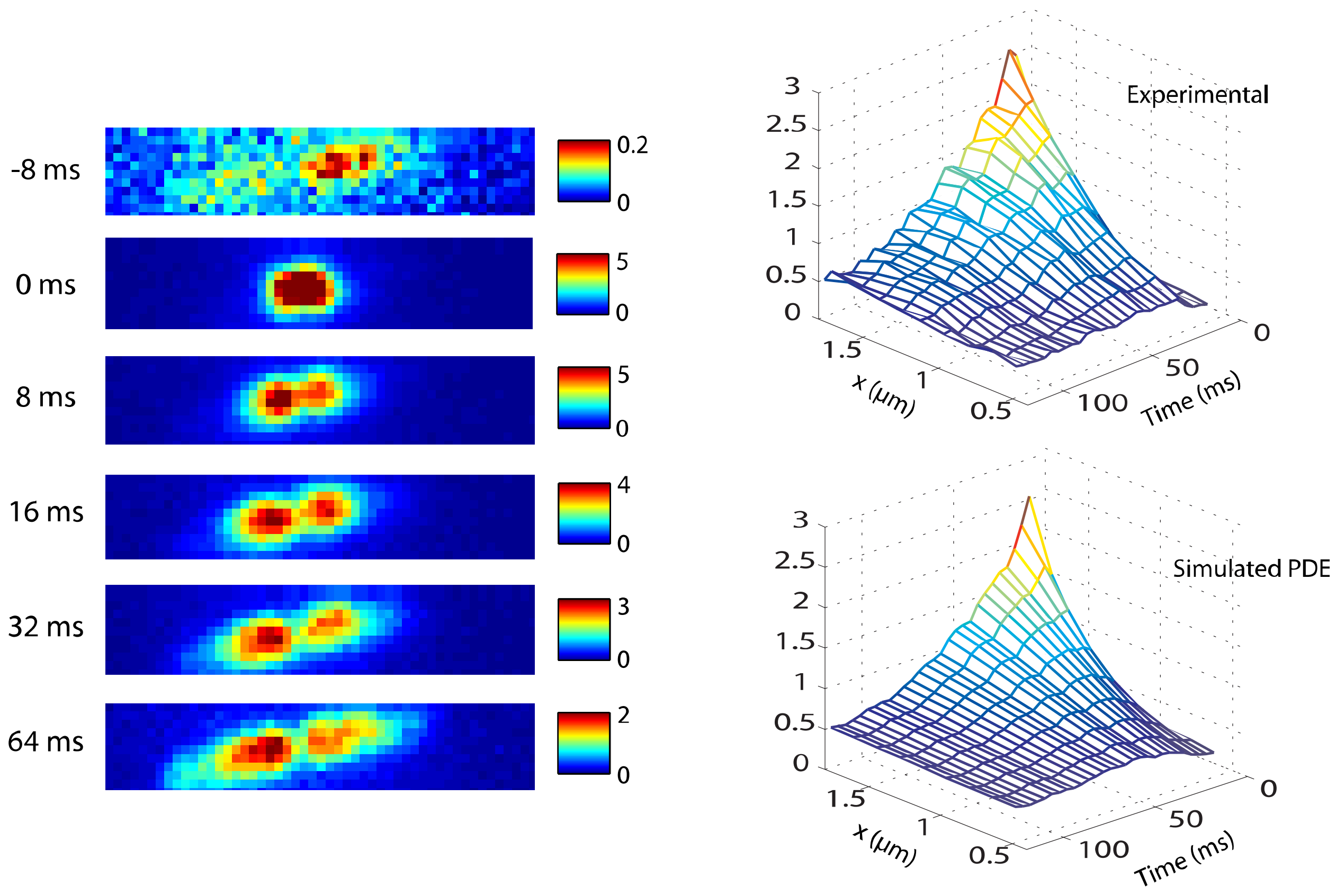

**Figure 6. Single-cell photoactivation ensemble experiments.**

Left panel: A large number of mEos2 molecules are activated in a diffraction-limited region at the interface of two *E. coli* cells at time zero. mEos2 spreads throughout the two cells over a time period of 64 ms.

Right top: The experimental fluorescence intensity values from the central part of the left cell are projected on the x-axis and plotted in the $(x, t)$-plane.

Right below: Solution of the 1D diffusion equation time-evolved with reflective boundaries corresponding to the lengths of the cells. The diffusion coefficient $D_{\text{micro}} = 11\ \mu\text{m}^2\ \text{s}^{-1}$ was obtained by fitting the experimental diffusion surface (*right top*).